\documentclass{aa}

\usepackage{natbib}
\bibpunct{(}{)}{;}{a}{}{,}

\usepackage{graphicx} 

\begin{document}

\title{On the origin of highest energy gamma-rays from Mkn~501}

\author{%
F.\ A. Aharonian\inst{1}, A.\ N. Timokhin\inst{1,2}, 
A.\ V. Plyasheshnikov\inst{1,3}
}

\offprints{F.\ A. Aharonian, \\ 
\email{Felix.Aharonian@mpi-hd.mpg.de}}

\institute{Max-Planck-Institut f\"ur Kernphysik, 
           Saupfercheckweg 1, Heidelberg, 69117, Germany
           \and 
           Sternberg Astronomical Institute,
           Universitetskij pr. 13, Moscow, 119899, Russia
           \and
           Altai State University, 
           Dimitrov Street 66, 656099 Barnaul, Russia}    

\authorrunning{F.A. Aharonian et al.}
\titlerunning{Highest energy gamma-rays from Mkn~501}

\date{Received / Accepted}

\abstract{%
The spectra of very high energy $\gamma$-radiation
from distant extragalactic objects suffer significant 
deformations during the passage of primary $\gamma$-rays
through the intergalactic medium. 
The recently reported fluxes of diffuse infrared background 
radiation indicate  that  we detect, most probably,  heavily  absorbed
TeV radiation   from  Mkn~421 and  Mkn~501.  This implies that the 
absorption-corrected  spectrum  of Mkn~501  may contain  
a sharp pile-up which  contradicts to the predictions  of  the conventional models of  
TeV blazars, and thus may leads  to the so-called ``IR background-TeV  gamma-ray crisis''.  
To overcome this difficulty,  
in this paper we propose two  independent hypotheses  assuming that 
(i) the TeV radiation from Mkn~501  has  a secondary origin, i.e. it is 
formed during the development of electron-photon cascades in the 
intergalactic medium initiated by primary $\gamma$-rays; (ii) 
the pile-up  in the source spectrum is a result of comptonization
(in deep Klein-Nishina regime)  of ambient optical radiation by 
an  ultrarelativistic conical cold outflow (jet) 
with bulk motion Lorentz factor $\Gamma_0 \geq 3  \times 10^7$. 
Within the uncertainties caused by the limited energy resolution 
of  spectral measurements,  the  observed TeV radiation of Mkn~501 formally  can be 
explained  by the   intergalactic cascade $\gamma$-rays, assuming 
however an extremely  low intergalactic magnetic field in the direction 
to the source  at the  level  of $\leq 10^{-18} \ \rm G$. 
We also demonstrate   that  the  ``bulk motion  comptonization''  
scenario   can quite  naturally  reproduce  the unusual spectral features in 
the absorption-corrected  TeV spectrum   of Mkn~501,  and briefly discuss the 
astrophysical implications  of this hypothesis.       
}

\maketitle

\section{Introduction}

The  Diffuse Extragalactic Background Radiation (DEBRA)  at infrared to 
ultraviolet  wavelengths carries crucial cosmological 
information about the galaxy formation epochs. 
It is believed that this radiation  basically consists  
of two emission components produced by stars and  
partly absorbed and re-emitted by dust 
during the entire history of evolution of galaxies.  
Consequently, two distinct bumps in the spectrum of 
red-shifted radiation at near infrared (NIR)  $\lambda \sim 1 - 2 \, \rm \mu m$
and far  infrared (FIR) $\lambda  \sim 100- 200 \, \rm \mu m$
wavelengths,  and a  mid infrared (MIR) ``valley''  between these bumps 
are expected  \citep[see e.g. ][]{dwek/::iv:1998, 
primack/:1999,  pei/:1999,  silk/devriendt:2000, 
malkan/stecker:2000, hauser/dwek:2001,franceschini/:2001}.

Direct measurements of  the Cosmic Infrared Background (CIB) 
radiation  contain large uncertainties 
because of heavy contamination caused by   
foregrounds of  different origin 
\citep[for review  see][]{hauser/dwek:2001}. 
Gamma-ray  astronomy offers a 
complementary approach to derive information about CIB. 
Although this method requires certain model assumptions about 
the primary (un-absorbed) spectrum of $\gamma$-rays, 
it has an adequate  potential for robust conclusions 
concerning the absolute flux and the spectrum of CIB. 
Moreover,  the study of angular and spectral properties of 
$\gamma$-radiation from predicted  giant electron-positron 
halos surrounding  powerful nonthermal  extragalactic objects 
like AGN and radiogalaxies   with known  redshifts  
\citep{aharonian/coppi/:1994} can provide a unique  tool
to ``measure''  unambiguously the broad-band 
spectrum and the absolute flux of CIB at different cosmological 
epochs,  and thus to probe the evolution of galaxies in past.    
 
The spectra of high energy ($E \geq 10 \, \rm GeV$)    
$\gamma$-radiation observed  from distant extragalactic 
objects suffer significant  deformation during  their  
passage through the intergalactic medium  due to  
interactions of  primary  $\gamma$-rays with CIB  
\citep{nikishov:1962,gould/schreder:1967,stecker/dejager/:1992}. 
The absorption features  in the $\gamma$-ray spectra 
depends on the flux of CIB, thus the study of such 
features from extragalactic objects with firmly
established distances  could yield important constraints on 
CIB. Strictly  speaking, this approach requires good 
understanding of the  {\em source} spectra of $\gamma$-rays 
from ensemble of sources located at different cosmological  
distances. Otherwise,  the conclusions based merely on $\gamma$-ray 
observations from a {\em single} source would be essentially 
model-dependent, and therefore would permit  
different interpretations  concerning both the intrinsic  
$\gamma$-ray spectrum, and the   flux of CIB.  

Presently we do face such an ambiguity, when trying to 
interpret the  multi-TeV $\gamma$-ray  emission of Mkn~501 observed   
during its  remarkably strong and long  flare  
in 1997 \citep{aharonian/::mkn501spectr:1999}.
The only definite   conclusion which can be drawn 
from these observations is that we see, 
most probably, significantly absorbed TeV radiation, 
especially at energies above 10 TeV, 
for which  the optical depth could be  as large as 10 
\citep{coppi/aharonian:1999,finkbeiner/davis/:2000,
protheroe/meyer:2000}. Moreover,  
a non-negligible absorption may take place 
already  at low, sub-TeV energies 
\citep{coppi/aharonian:1999, guy/:2000}. 
It should be noticed that  the analysis of the intergalactic 
absorption at sub-TeV and multi-TeV parts of the 
spectrum of Mkn~501  leads to two essentially  
different conclusions.  The  
{\it absorption-corrected} $\gamma$-ray spectrum 
at low energies,  based on the CIB fluxes reported  
at 2.2 and  3.5 $\mu \rm m$ 
\citep{dwek/arendt:1998,gorjian/:2000,wright:2001} 
and on the current theoretical predictions for 
NIR \citep[see e.g.][]{primack/:2001}, is in a general 
agreement with the Synchrotron-self-Compton (SSC)  
model of X- and TeV radiation of Mkn~501 
\citep[see e.g.][]{guy/:2000, krawczynski/coppi/:2000, primack/:2001}.
On the other hand,  
the corrections to the  $\gamma$-ray  spectrum at energies above 
10 TeV based on the unexpectedly large CIB fluxes detected by COBE 
at 140 and  240   $\mu \rm m$  
\citep{hauser/::dirbe:1998, schlegel/:1998, lagache/:1999}  
result in an  ``unreasonable''  source spectrum.  Namely it implies   
a source spectrum which  sharply {\it curves up} above 10 TeV, 
unless  we assume that the CIB flux at MIR between 
10 and 50 $\mu \rm m$ is quite low (a few $\rm nW/m^2 sr$),  
and at longer wavelengths it 
increases rapidly ($\nu F_\nu \propto \lambda^{s}$   
with $s \geq  2$) in order to match the COBE points 
\citep{aharonian/::mkn501spectr:1999,coppi/aharonian:1999,
renault/:2001}.
Even so, the large DEBRA fluxes at 140 and  250   $\mu \rm m$  
imply very flat ``reconstructed''  source $\gamma$-ray   
spectral energy distributions  (SED),    
$\nu S_\nu = E^2  \mbox{d}N/\mbox{d}E=
E^{2-\alpha}$ with a photon index $\alpha \leq 2$.
Such flat source spectra  extending beyond 10 TeV 
require,  within  SSC models,  rather unconventional jet
parameters, namely very large Doppler factors and very small 
magnetic fields  (H.~Krawczynski and P.~Coppi, private communication;
J.~Kirk, private communication).

A real  trouble   arises, however, when we take 
into account the recent claims about detection of 
CIB flux at 100  $\mu \rm m$  
\citep{lagache/:1999,finkbeiner/davis/:2000},
 and especially at 60  
$\mu \rm m$  \citep{finkbeiner/davis/:2000}. 
If we refer the reported fluxes to the  truly diffuse 
extragalactic  background radiation, then a little  
room would be left  for speculations concerning the spectral shape 
of CIB in order to  prevent the  ``unreasonable''  
$\gamma$-ray source spectrum. This implies that we should accept 
the existence of a sharp pile-up in the spectrum of TeV radiation. 
Motivated by such a non-standard spectral shape of TeV-radiation, 
recently several extreme  assumptions have been made in order to 
overcome the ``IR background -- TeV gamma-ray crisis''
\citep{protheroe/meyer:2000}. 
In particular, \citet{harwit/:1999}
suggested  an interesting hypothesis that the HEGRA highest energy events 
are due to Bose-Einstein condensations interacting  with the air 
atmosphere, and proposed a test to  inspect this hypothesis by 
searching for peculiar features of showers detected by HEGRA 
in the direction of Mkn~501. Subsequently, 
the HEGRA collaboration has demonstrated 
\citep{aharonian/::bose:2000}  
that  the detected shower characteristics are in fact in good 
agreement with the predictions  for the   
events initiated by {\it ordinary}  $\gamma$-rays. 
Another, even more dramatic hypothesis
-- violation of the  Lorentz invariance -- has been 
proposed by several authors 
\citep[see e.g.][]{coleman/glashow:1999, kifune:1999, kluzniak:1999, 
aloisio/:2000, amelino-camelia/piran:2001, protheroe/meyer:2000} to solve this
problem.   
We may add to the list of  ``exotic''  solutions
of the ``IR background -- TeV gamma-ray crisis'' 
a less dramatic, in our view, hypothesis, assuming that 
Mkn~501 is located at a distance significantly less than 100~Mpc  
--  good news for the advocates of non-cosmological 
origin of some of  AGN and quasars 
\citep[see e.g.][]{hoyle/burbidge:1996, arp/:1997}.

Although very fascinating, it seems to us too premature  
to invoke such dramatic revisions  of essentials of 
modern physics and astrophysics.  The nature of the FIR 
isotropic emission detected by COBE is  not yet firmly established,
and  it is quite possible that the bulk of the reported flux,
especially below 100 $\mu \rm m$, is a result of superposition of
different local backgrounds.
Needless to say, that this would be the simplest solution of the 
problem  \citep{stecker:2000, renault/:2001}. 
In  this paper we adopt, however,  that  the reported  
FIR fluxes  have universal (extragalactic) origin. 
This implies that we  adopt the existence of a pronounced 
pile-up in the $\gamma$-ray spectrum of Mkn~501
above 10 TeV, but  try to find  an  explanation of this  
spectral feature within the framework 
of new but  yet  conventional astrophysical scenarios. 
In this paper we propose and study  two  potential 
ways to overcome the  ``IR background -- TeV gamma-ray crisis'':

(1)  TeV $\gamma$-rays  from  Mkn~501 are not direct 
representatives of  primary radiation of the source, but 
have a secondary origin, i.e. they are formed 
during the development of 
high energy electron-photon cascades in the intergalactic medium
initiated by interactions of primary $\gamma$-rays with 
diffuse extragalactic photons. 
This hypothesis implies 
a rather extreme  assumption  concerning the strength of the  
intergalactic magnetic field on $\geq 1 \, \rm Mpc$,  
$B \leq 10^{-18} \, \rm G$. 

(2) The  ``reconstructed'' source spectrum of Mkn~501
with a flat (almost constant) SED below 10 TeV, and 
a pile-up beyond  10 TeV, is partly  
or entirely produced by monoenergetic ultrarelativistic beam of 
electrons due to the  inverse Compton scattering  in  deep 
Klein-Nishina regime. 
In order to avoid 
significant radiative  (synchrotron and Compton) losses,
which otherwise would result in an equilibrium,  
$E_{\rm e}^{-2}$ type differential spectrum of electrons,  
we assume that such a beam of electrons in fact is a cold, 
conical kinetic-energy dominated  ultrarelativistic wind 
with the bulk Lorentz factor  $\Gamma_0 \sim 4 \times 10^{7}$ 
formed  beyond the accretion disk of the central black hole.  
Apparently this  hypothesis implies non-acceleration origin of 
highest energy $\gamma$-rays detected from Mkn~501.

\section{Absorption of gamma-rays in CIB}

\begin{figure}[tb]
\begin{center}
\includegraphics[width=0.9\linewidth]{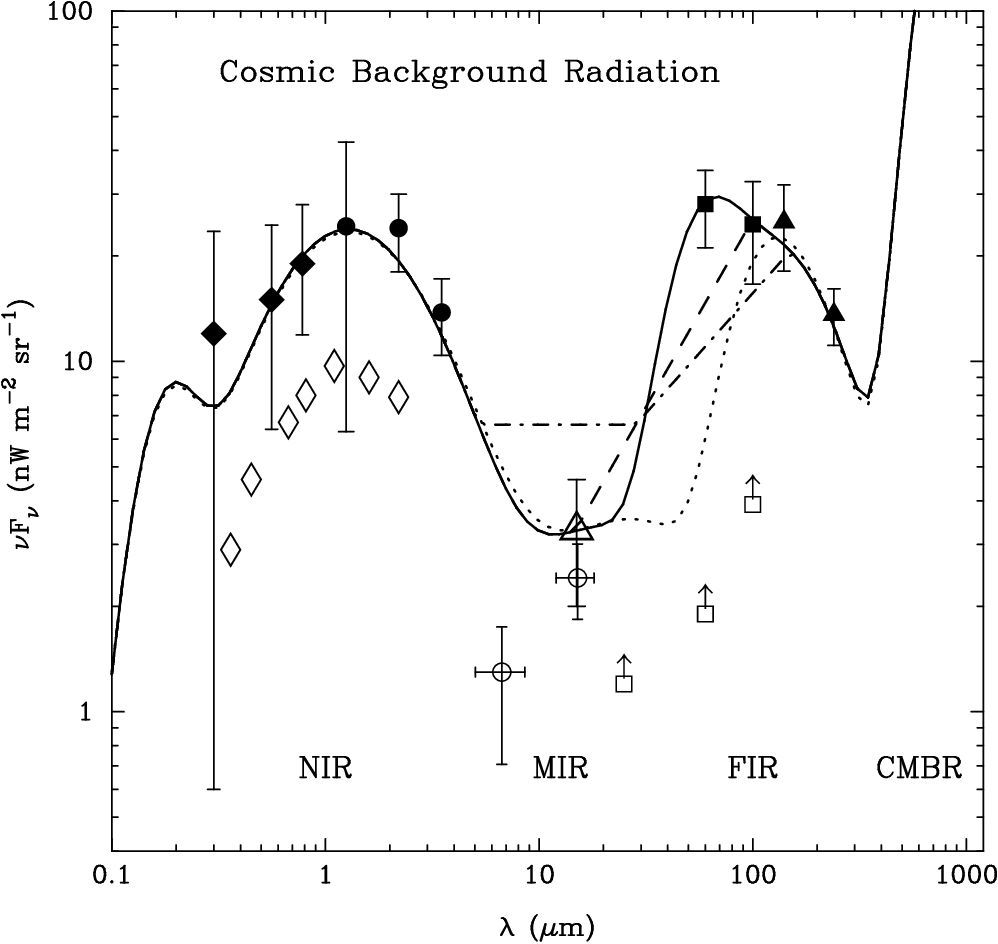}
\caption{Cosmic background  radiation. 
The reported fluxes are shown with filled symbols:
\citet{bernstein:1999} -- diamonds,  
\citet{wright:2001} -- circles, 
\citet{finkbeiner/davis/:2000} -- squares, 
\citet{hauser/::dirbe:1998} -- triangles. 
The low limits are shown by open symbols:
\citet{pozzetti/:1998} -- diamonds, 
\citet{biviano/:2000} -- triangle,
\citet{franceschini/:2001} -- circles,
\citet{hacking/soifer:1991} -- squares,
The  CIB models are shown by   
solid line -- Model I, by dotted line -- Model II, 
by dashed line --  Model III,
by dot-dashed line --  Model IV (for details see the text)
}
\label{SED_CBR}
\end{center}
\end{figure}

The reported  fluxes and flux upper/low limits of CIB 
from optical/UV to far IR wavelengths are shown in 
Fig.~\ref{SED_CBR}. The  reliability and the implications 
of these measurements 
are discussed in  the recent review article by 
\citet{hauser/dwek:2001}.
%
%
The level of the spectral energy distribution (SED) of
CIB at optical/NIR wavelengths  
with  the  ``best guess estimate'' between 20 and 50
$\rm nW/m^2 sr$  is comparable with the overall energy 
flux of FIR of about 40-50 $\rm nW/m^2 sr$ 
\citep{pozzetti/madau:2000}.
 This indicates that
an essential part of the energy radiated by stars is absorbed and 
re-emitted by dust in a form of thermal sub-mm emission.
Currently the information  at mid-infrared 
wavelengths is very limited. The  only 
available measurement  at  6 and 15 $\mu$m on Fig.~\ref{SED_CBR} 
derived from the ISOCAM source counts \citep{franceschini/:2001}
shown in  Fig.~\ref{SED_CBR} 
should be  treated as a lower limit. 
In Fig.~\ref{SED_CBR} we show also a slightly higher flux estimate
at 15 $\mu$m reported by \citet{biviano/:2000}.  
Therefore the flux estimate 
at the level of $\simeq 2-3  \rm nW/m^2 sr$ 
\citep{franceschini/:2001}  as well as the lower limits 
based on the IRAS counts at 
25-100 $\mu \rm m$ \citep{hacking/soifer:1991} 
do not allow  firm conclusions about the depth 
of the MIR ``valley''   dominated  by radiation of the 
warm dust component.  Consequently, it does not provide 
sufficient  information  for definite predictions 
regarding the slope of the spectrum  in the most crucial 
(from the point of view of absorption of 
$\geq 10 \, \rm TeV$  $\gamma$-rays)  
MIR-to-FIR transition region. 

As long as the available measurements of CIB do not allow 
a quantitative study of the effect of absorption of $\gamma$-rays 
in the intergalactic medium, we can rely only on model predictions 
or on the ``best guess''  shape of the CIB spectrum.
In this regard we notice that the reported high FIR fluxes 
present a common problem  for all  current CIB  models. 
Therefore,  if one adopts that the reported FIR fluxes  
have truly diffuse extragalactic origin, 
an essential revision of the CIB models is needed in order to 
match the data.  Such an attempt has been made recently by 
\citet{primack/:2001} who showed that 
their semi-analytical approach, with a 
reasonable  adjustment of  some model parameters,
and  using the empirical dust emission 
templates of \citet{dwek/::iv:1998},
can match  the reported FIR fluxes. 

\begin{figure}[tb]
\begin{center}
\includegraphics[width=0.9\linewidth]{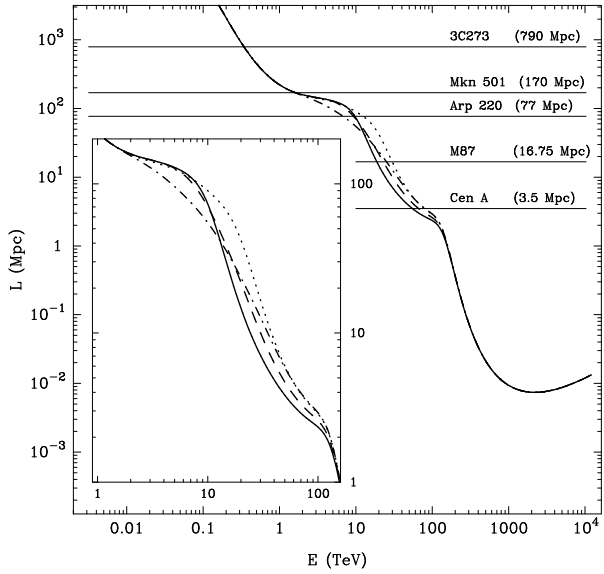}
\caption{Mean free path of $\gamma$-rays for 4 different models
of the CIB spectrum:  solid line -- Model I, dotted line -- Model II, 
dashed line --  Model III, dot-dashed line -- Model IV.  
The horizontal lines indicate  the distances to 
Cen A, M87,  Mkn~501, and  3C 273 ($H_0=60 \ \rm  km/s \ Mpc$).  
}
\label{Lambda}
\end{center}
\end{figure}

\begin{figure}[tb]
\begin{center}
\includegraphics[width=0.9\linewidth]{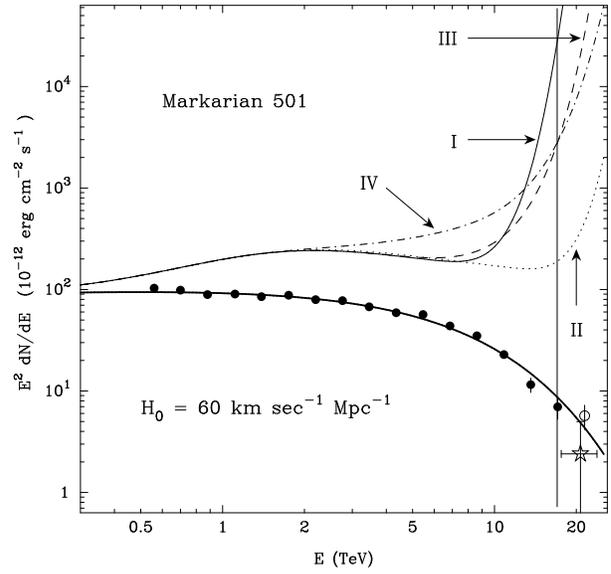}
\caption{Spectral Energy Distribution (SED) of Mkn~501.
The experimental points (filled circles) correspond to the 
time-averaged spectrum of Mkn~501 during the flare in 1997
\citep{aharonian/::mkn501spectr:1999}. 
The heavy line corresponds to the fit to this data
in the form of Eq.~(\ref{eq:hegra_501}).
The star correspond to the flux in the highest energy bin 
around 21 TeV obtained after the reanalysis of the same data set 
but with improved energy resolution 
\citep{aharonian/::mkn501newspectr:2001}. 
The vertical line at 17 TeV indicates the edge of the spectrum 
of Mkn~501 measured by HEGRA with high statistical significance.  
The solid, dotted, dashed  and dot-dashed lines represent
the reconstructed  (absorption-corrected) spectra of $\gamma$-rays  for  
the CIB Models I,II, III and IV, respectively ($H_0=60 \ \rm km/s \ Mpc$).
}
\label{mkn501_60}
\end{center}
\end{figure}

In Fig.~\ref{SED_CBR} we show several  model spectra  of CIB.  
Since we are interested,  first of all,  in the cosmic background fluxes 
at MIR and FIR, at shorter wavelengths  we adopt a common for all models 
approximation which matches the reported  optical and 
NIR fluxes.   The  smooth template of the CIB spectrum 
in the most principal  MIR-to-FIR transition region
shown by dashed line (hereafter Model III)  in Fig.~\ref{SED_CBR},
fits the  reported  fluxes including, within $2\sigma$ uncertainty, the 
60 $\mu \rm m$ point.    Actually this idealized template is quite  
similar to the CIB spectrum shown in Fig.~1 of 
\citet{franceschini/:2001}  based on their  
reference model for IR  galaxy evolution, as well as to 
some of the recent theoretical models of  \citet{primack/:2001}.  
It has a rather flat shape in the MIR-to-FIR transition region, 
$\nu F_\nu \propto  \lambda^s$ with $s \leq 1$. 
This results in  short mean free paths  of $\gamma$-rays 
above 10 TeV  ($L \leq 50 \, \rm Mpc$; see Fig.~\ref{Lambda}), 
and consequently in a  pile-up in the  
reconstructed $\gamma$-ray source spectrum 
(see Fig.~\ref{mkn501_60}) defined  as:
\begin{equation} 
J_0(E)=J_{\rm obs}(E) \ \exp{[\tau(E)]} \, ,
\label{eq:gsource}
\end{equation}
where  
$J_{\rm obs}$ is the {\em observed}  $\gamma$-ray spectrum,
$\tau(E)=d/L(E)$ is the intergalactic optical depth, and 
$d$ is the distance to the source.
%
%

If we adopt that the reported FIR 
fluxes correctly describe the level of truly
diffuse background radiation,  only 
two ways are  left for reduction   of the 
effect of attenuation of $\geq 10 \ \rm TeV$ $\gamma$-rays:  
(i) an  {\it ad hoc} assumption of 
the CIB flux at wavelengths between 10 and  60  $\mu m$ 
at the marginally acceptable (i.e.  the ISOCAM low-limit) level, 
but with  very rapid rise beyond 60 $\mu m$ in order to match 
the  reported fluxes at FIR, and (ii)  adopting  for  
the Hubble constant $H_0 \simeq 100 \ \rm km/s \ Mpc $, i.e. 
assuming the smallest  possible distance to Mkn~501  
$d=c z/H_0 =  102 \, \rm Mpc$ ($z=0.034$). 

In Fig.~\ref{SED_CBR} we show 2 other model spectra 
of CIB (solid line -- Model  I, dotted line -- Model  II)
which fit the data at NIR and FIR, but at the 
same time allow minimum intergalactic $\gamma$-ray absorption
at $E \geq 10 \ \rm TeV$,   
because both spectra  are forced to be at the lowest possible 
level at MIR-to-FIR transition region set by the ISOCAM lower 
limit at 15 $\mu \rm m$.  

%
%

The CIB fluxes above 140 $\mu \rm m$  are approximated by the 
function suggested by  \citet{fixsen/:1998}.  Below 140 $\mu \rm m$ 
for both  I and II Models   we assume the same spectral 
shape  described by Planckian  distribution.    
Such spectra  should be considered as a lower 
limit, because in the most
critical MIR-FIR transition region near $60\mu$m 
it is essentially the Wien part
of the blackbody spectrum -- 
the steepest (physically justified) continuous spectrum.
The Model II marginally  agrees with the 100 $\mu \rm m$  point but 
underestimates the flux at 60 $\mu \rm m$ by a factor of 5 
compared  with the flux reported by \citet{finkbeiner/davis/:2000}.
This model  significantly suppresses the  photon density 
between 50 and 100 $\mu \rm m$, and correspondingly allows 
larger mean free paths for $\gamma$-rays with energy more 
than several TeV  (Fig.~\ref{Lambda}). Such a SED of 
CIB results in almost  $E^{-2}$ type ``reconstructed''  spectrum 
of $\gamma$-rays from Mkn~501 at energies between approximately 
2 and  20 TeV (Fig.~\ref{mkn501_60}). 
This has a simple explanation.  For the constant 
SED of CIB (the flat part of the spectrum 2 in Fig.~\ref{SED_CBR}), 
$n_{\rm CBR}(\epsilon) \propto \epsilon^{-2}$,
the photon-photon optical depth is proportional to $E$ 
with an absolute value for  Mkn~501   
$\tau(E) \approx 0.16 (E/1 \ \rm TeV)$  \citep{aharonian/::mkn501spectr:1999}.
Therefore, in the absorption-corrected (source)  spectrum of 
$\gamma$-rays given by Eq.~(\ref{eq:gsource}),
the optical depth $\tau$ almost compensates  
the exponential term of the {\rm observed} 
spectrum of Mkn~501%
\footnote{
Note that somewhat different CIB models with more 
complex spectral shapes can also lead to  flat,  $E^{-2}$ type 
power-law absorption-corrected $\gamma$-ray spectra  
\citep[see e.g.][]{coppi/aharonian:1999, konopelko/:1999, guy/:2000}}, 
which from 0.5 TeV to 20 TeV can be presented in a 
simple ``power-law with exponential cutoff'' 
form \citep{aharonian/::mkn501spectr:1999}  
\begin{equation}
J_{\rm obs}(E)   \propto E^{-1.92} \exp{(-E/6.2 \ \rm TeV)} \ . 
\label{eq:hegra_501}
\end{equation}

Recently, the HEGRA collaboration published the results 
of reanalysis of the spectrum of Mkn~501
\citep{aharonian/::mkn501newspectr:2001} 
using  an improved method of shower energy reconstruction with 
10 to 12 per cent energy resolution \citep{hofmann:2000}. 
Apparently the improved energy resolution is more relevant for derivation of the
shapes of steep $\gamma$-ray spectra, e.g. spectra with exponential or sharper
cutoffs.  The new HEGRA spectral analysis  confirmed the  results of the
previous study  up to energy 17 TeV, but   shows a steeper 
spectrum beyond that energy 
\citep{aharonian/::mkn501newspectr:2001}. 
Therefore in Fig.~\ref{SED_CBR} we show only one point of  
the new  analysis - the flux in the highest energy bin 
around 21 TeV (star) which lies significantly below
the flux estimate   of the previous study (open circle). 
We should notice, however, that the points above 
17 TeV  in both papers do not have   
high statistical significance anyway,  
therefore below we limit our 
study only by the energy region up to  17 TeV (i.e.
by the region on the left side of the vertical line shown in  
Fig.~\ref{mkn501_60}). 
In this energy interval the energy spectra obtained  
by new and old methods  are perfectly  described 
\citep{aharonian/::mkn501newspectr:2001}
by  Eq.~(\ref{eq:hegra_501}).

Apparently, for conventional  $\gamma$-ray production 
mechanisms  the $E^{-2}$  type $\gamma$-ray source 
spectrum sounds more comfortable than 
the spectrum  containing  a sharp pile-up.
Therefore, the Model II in Fig.~\ref{SED_CBR} can be treated as 
the ``most favorable''  CIB spectrum, although its flat shape 
in the broad MIR-to-FIR transition region with very fast increase 
beyond  $60 \ \mu \rm m$  hardly match the  current theoretical 
descriptions of CIB, as well as  the  FIR spectra observed from 
nearby galaxies.  Such a comparison is of special interest, especially 
because  the bulk of the integrated extragalactic 
background radiation  has  been generated  presumably at $z<1$ 
\citep{harwit:1999,franceschini/:2001}. In particular, 
the shape of the Model II  contradicts to the 
average SED of  the ISOCAM sources which not only 
contribute a dominant fraction to  CIB at MIR, but  
likely are major contributors at longer wavelengths as well 
\citep{franceschini/:2001}. 
   
\begin{figure}[tb]
\begin{center}
\includegraphics[width=0.9\linewidth]{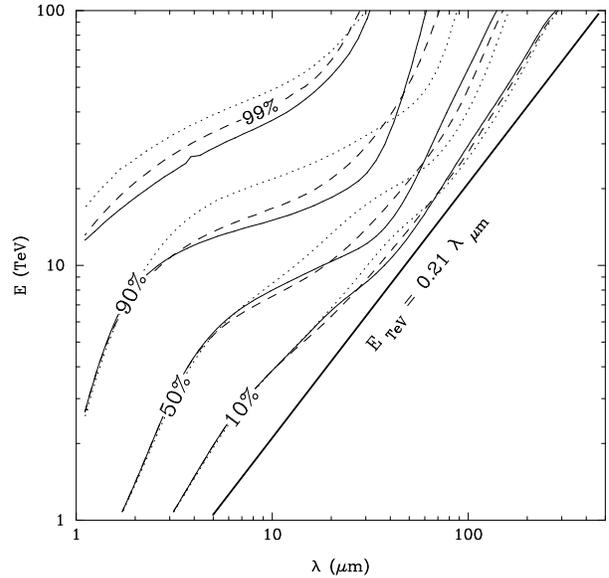}
\caption{The contour map of the function $\kappa(\lambda,E)$.
The solid, dotted and dashed lines show the levels
of $\kappa(\lambda,E)$ for  
the CIB Models I, II, and III, respectively.
The heavy solid line represents the threshold 
of $\gamma\gamma$ pair production.} 
\label{kappa}
\end{center}
\end{figure}

All current CIB models  have a problem to accommodate  
the reported flux at 60 $\mu \rm m$. If this point, however, is the   
representative of  the truly diffuse flux, we must assume a CIB spectrum 
close to the 
Model - I shown in  Fig.~\ref{SED_CBR} by the  solid line.
This  Model   assumes a very steep slope
between 30 and 60 $\mu \rm m$. A
steeper  spectrum (i.e. steeper  than the Wien tail of the 
black-body radiation) in this narrow wavelength band 
hardly could be physically justified.
Moreover, already 60 $\mu \rm m$ photons have sufficient energy for 
effective  interaction with $\geq 10 \ \rm TeV$ $\gamma$-rays, 
therefore we cannot suppress anymore the severe $\gamma$-ray 
absorption by speculating about   the spectral shape of CIB (see  
Fig.~\ref{kappa} and the related discussion below).

Finally,  in  Fig.~\ref{SED_CBR} we show one more possible model spectrum 
of CIB (dot-dashed curve -- Model IV) which assumes significantly,  by a factor of 2.5,
higher flux at 15 $\mu \rm m$ compared to  the reported 
ISOCAM lower,
and smoothly passes through the low edges of the error bars 
of reported fluxes at 
100 and 140 $\mu \rm m$.  Surprisingly such a  
high at MIR spectrum does not result in an unusual 
$\gamma$-ray spectrum as long as it concerns the 
energy region below 10 TeV. Namely,  at these energies we obtain an  
almost single  hard power-law  $\gamma$-ray source spectrum  
with photon  index less than 2.  Above 10 TeV we again observe 
a pile-up which however in this case is less pronounced than in the case of 
Models I and III.

%
%

A $\gamma$-ray photon with energy 
$E$ propagating trough isotropic photon field can interact, 
via electron-positron pair production, with ambient photons of energy
$\epsilon \ge \epsilon_{\rm th} = (m_{\rm e} c^2)^2/E\simeq
0.26(E_\gamma/1\mbox{TeV})^{-1} \ \rm eV$ or of 
wavelength $\lambda = 4.8 (E_\gamma/1\mbox{TeV}) \ \mu \rm m$.
The  latter relation is represented  in Fig.~\ref{kappa} by the solid line. 
The cross-section of 
$\gamma\gamma$ interactions  averaged over the 
directions of background photons peaks at
$\epsilon_{\rm max}\simeq 3.5 \epsilon_{\rm thresh}$ with
$\sigma^{\rm max}_{\gamma\gamma} \simeq (1/4) \sigma_{\rm T}$
\citep[see, e.g.][]{vassiliev:2000} 

Therefore,  even for a broad, e.g.  power-law spectrum of 
background radiation,  the most contribution to the optical 
depth $\tau$ comes from a narrow spectral band of CIB within 
$\epsilon_{\rm max}\pm \Delta \epsilon$ with 
$\Delta \epsilon \sim 1/2 \epsilon_{\rm max}$.
For typical spectra  of CIB with
two distinct NIR and FIR bumps and a 
MIR valley,  the ``$\gamma$-ray energy-CIB photon wavelength''
relation is more complicated, namely
the relative contributions of different spectral 
intervals of CIB to the optical depth $\tau_0$ 
significantly depend on 
the energy of the primary $\gamma$-ray photon.   
It is convenient to describe this dependence by the ratio 
$\kappa(E,\lambda)=\tau(E, \lambda)/\tau_0(E)$,
where 
\begin{equation}
\tau(E,\lambda) = \int_{\lambda}^{\lambda_{\rm th}}
\sigma_{\gamma \gamma}(E_{\gamma},\lambda) 
n_{\rm CBR}(\lambda)\, 
{\rm d} \lambda \ ,
\label{eq:tau}
\end{equation}   
and $\tau_0=\tau(E_\gamma, \lambda=0)$ is  the 
optical depth integrated over the entire spectral range of 
CIB $n_{\rm CIB}(\lambda)$  above the threshold    
$\lambda_{\rm th} \simeq 4.8 \ (E_\gamma/1 \ \rm TeV) \ \mu m$.

In Fig.~\ref{kappa} the contour map of the function
$\kappa(E, \lambda)$ is shown for 3 different models of CIB
in the spectral range of $\gamma$-rays from 1 to 
100 TeV,   and for the CIB photons 
from 1 to 500 $\mu \rm m$. For the given energy of 
$\gamma$-ray photon $E_0$, the spectral region 
of CIB responsible for the fraction
$\xi$  of the total optical depth $\tau_0$ is 
$[\lambda_\xi, \lambda_{\rm th}]$, where 
$\lambda_\xi$ and $\lambda_{\rm th}$   are the 
abscissas of the points where the horizontal line $E=E_0$ 
intersects the corresponding level curve,  and the threshold 
line, respectively. In Fig.~\ref{kappa} four levels
for $\xi=10, 50, 90$, and $99 \%$, are shown.
By definition, $\xi=0$ corresponds to the threshold boundary 
$E_{\rm TeV}=0.21 \ \lambda_\mu$. 
It is seen that at $E_0=17 \ \rm TeV$  
approximately 50 per cent of the  
total optical depth $\tau_0$ is contributed by 
background photons with wavelengths longer than  
$50 \mu \rm m$. Therefore even the sharp 
cutoff of the CIB flux below $50 \mu \rm m$  cannot prevent 
the large optical depth of 17 TeV $\gamma$-rays, as far as we 
accept that  the $60 \mu \rm m$ flux has a 
cosmological origin, and reflects  the level of CIB 
at these wavelengths. This is the case of Model I for which 
the free path of 17 TeV $\gamma$-rays is about 20 Mpc,
and therefore for the distance to the source of 170 Mpc 
($H_0=60 \ \rm km/s \ Mpc$)   
the  optical depth $\tau_0 \approx 8.5$. The corresponding 
absorption factor is so large 
($\exp(-8.5) \simeq  2  \times 10^{-4}$) that  
the sharp file-up in the absorption-corrected 
spectrum becomes unavoidable. 

Since the exponent in the absorption  factor is very large, even a 
relatively  small reduction of the optical depth $\tau_0$ 
may help  to  reduce  significantly  the effect of  
$\gamma$-ray attenuation. 
In particular, the Model III,  which has a more 
realistic spectral shape at mid infrared wavelengths, and 
fits the FIR data above 100 $\mu \rm m$,  allows a bit larger 
(25 per cent or so) free paths at highest energy $\gamma$-rays.  
Although this  makes the pile-up less pronounced,  but still does not   
eliminate it altogether. Only the CIB Model II gives a comfortable, 
$E^{-2}$ type spectrum. 

We may reduce the absorption effect furthermore,  
assuming a larger, although presently less favored value 
for the Hubble constant, $H_0=100 \, \rm km/s \ Mpc$. 
The latter makes smoother the pile-up for the Model I, 
and almost removes  it for the Models III and IV. 
Therefore we may conclude that it is possible, in principle,  
to avoid the  ``disturbing''   sharp turn-up in the intrinsic
$\gamma$-ray spectrum of Mkn~501, if we adopt
for CIB a model like the Model III, and assume a very large 
value for the Hubble constant. 

Below we discuss, however,  possible solutions  which 
allow accommodation of  both higher 
CIB fluxes and more realistic value for the Hubble constant.

\section{The effect of cascading in the intergalactic photon fields} 
Generally, the propagation of high energy $\gamma$-rays
trough a low frequency photon field  cannot be reduced to 
the simple effect of $\gamma\gamma$ absorption. 
When a $\gamma$-ray is absorbed 
its energy in fact is not lost. The  secondary electrons
and positrons create new $\gamma$-rays via inverse 
Compton scattering;  the second 
generation $\gamma$-rays
produce new ($e^+,e^-$) pairs, thus an electromagnetic  
cascade develops.  Actually,  in the  
intergalactic space this process is  inevitable, and it may significantly
contribute   to  the  isotropic (extragalactic) $\gamma$-ray 
background  radiation 
\citep[][]{protheroe/stanev:1996, coppi/aharonian:1997}.
Note that in the energy region of interest the $\gamma$-rays 
and electrons interact with photon fields of different origin.  
Although the energy density of the 2.7 K cosmic microwave 
background radiation  (CMBR) well   exceeds the  
density of other (``starlight'' and ``dust'') 
components of  the cosmic background radiation  (see Fig. \ref{SED_CBR}), 
because of the kinematic threshold of the  reaction 
$\gamma \gamma \rightarrow  e^+e^-$   the $\gamma$-rays with 
energy less than several hundred  TeV interact mostly  with the infrared  
and optical background photons.   The inverse Compton scattering of 
electrons does not have kinematic  threshold, therefore the electrons 
interact predominantly  with  much denser CMBR.  

For intrinsic $\gamma$-ray spectra harder 
than $E^{-2}$ extending to energies $E \geq 100$  TeV , 
the cascade spectrum may 
well dominate over the primary $\gamma$-ray spectrum.  
On the other hand, the cascade spectrum typically has a standard
shape which slightly depends on the primary $\gamma$-ray spectrum. 
This makes rather attractive the idea of interpretation of one of the 
most remarkable features of the  TeV radiation 
of Mkn~501, namely  the  surprisingly  stable spectral shape 
of the source in high state, despite 
dramatic variation of the  absolute flux on timescales less 
than several hours   \citep{aharonian/::mkn501spectr:1999, 
aharonian/::mkn501timing:1999}. 
Moreover,  as it was noticed by 
~\citet{aharonian/::mkn501spectr:1999}, 
the shape  of the  TeV spectrum of Mkn~501 
given by   Eq. (\ref{eq:hegra_501})  resembles  
the $\gamma$-ray spectrum formed during the 
cascade development in the photon field.
Therefore,  the quantitative 
study of this effect presents a definite interest. 

The interpretation of 
the observed TeV emission
from Mkn~501 in terms of the intergalactic cascade radiation requires 
an extremely  low intergalactic magnetic field.
Indeed, for the field exceeding $10^{-12} \, \rm G$,
the $\gamma$-rays of the cascade origin should be observed in a form 
of an extended emission from a giant isotropic pair halo with an 
angular radius  more than several degree 
\citep{aharonian/coppi/:1994}.  
Both the detected  angular size and the time variability of the TeV 
radiation from Mkn~501 excludes such a possibility 
\citep{aharonian/::mkn501timing:1999, aharonian/::mkn501halo:2001}.
 For lower  magnetic fields,   the cascade 
$\gamma$-rays  penetrate almost on a straight line,  thus the first 
argument based on the  detected angular size becomes less stringent. 
Note however that  for the distance to the source $d \geq 100 \, \rm Mpc$
even tiny deflections of the secondary (cascade) electrons
by the intergalactic magnetic field would lead  to non-negligible 
time delays of arriving $\gamma$-rays 
\citep{plaga:1995, kronberg:1995}:   
\begin{equation} 
\Delta t_{\rm B} \simeq 
10 (d/170\mbox{Mpc})(E/1\mbox{TeV})^{-2} (B/10^{-18}\mbox{G})^2
\mbox{ h} 
\label{eq:igmf}
\end{equation}
This would smear out the time variation of  TeV fluxes observed  
on timescales less than  several hours 
\citep{aharonian/::mkn501timing:1999}, unless
$B \leq 10^{-18} \, \rm  G$%
\footnote{Actually,
due to non-zero ($\sim m_\mathrm{e} c^2/E$) 
emission angles of  the secondary products
in the $\gamma \gamma \rightarrow e^{+}e^{-}$ and
$e\gamma \rightarrow e \gamma^{\prime}$ reactions,  
we should expect a non-negligible 
broadening,  and consequently  time delays of the 
cascade radiation  even at  the
absence of magnetic field \citep{cheng/cheng:1996}.}.

For such  a low intergalactic magnetic field,   we have studied 
the spectral  properties of  electromagnetic cascades 
generated  by the  primary multi-TeV  radiation of Mkn~501 
in  the intergalactic medium,  using  a fast numerical 
method based on the  solution of adjoint cascade equations 
\citep{uchaikin/:1988}.
Two processes have been taken into account:  
electron-positron pair production 
and inverse  Compton scattering.  The details of calculations will be 
published elsewhere.

%
\begin{figure}[tb]
\begin{center}
\includegraphics[width=0.9\linewidth]{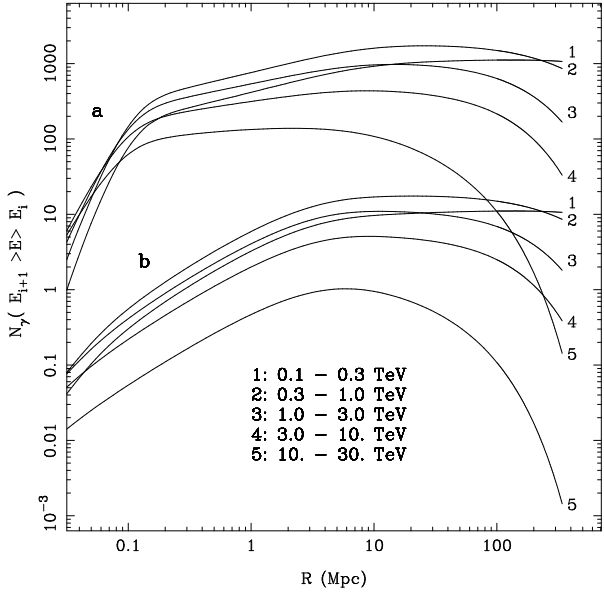}
\caption{The cascade development initiated by primary 
monoenergetic $\gamma$-rays  with energy 
(a) $E_0=10^4$ TeV  and (b) $E_0=10^2$ TeV for the CIB 
Model I. The curves show the number of cascade $\gamma$-rays 
in different energy intervals as a function of the penetration depth. 
} 
\label{cas_dev1}
\end{center}
\end{figure}  

\begin{figure}[tb]
\begin{center}
\includegraphics[width=0.9\linewidth]{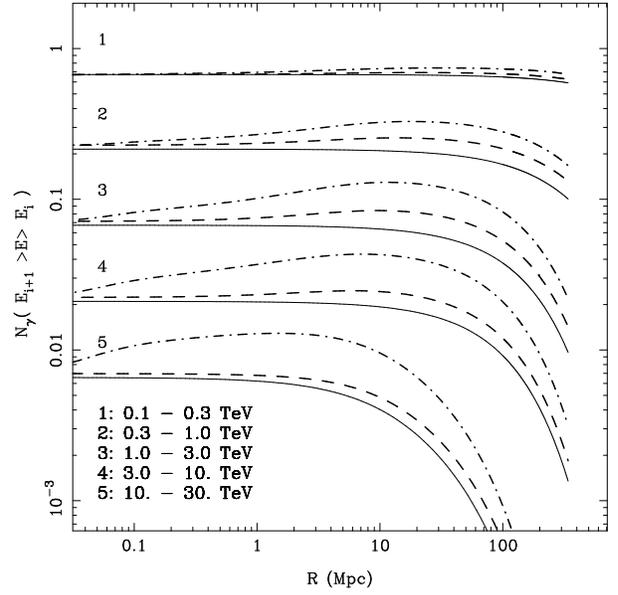}
\caption{The same as in Fig.~\ref{cas_dev1}, but 
for the power-law spectrum of primary $\gamma$-rays, 
$J_0(E) \propto E^{-2}$ extending to $E_{\rm max}=10^2$ TeV 
(dashed lines) and $E_{\rm max}=10^4$ TeV (dot-dashed lines). 
The solid lines show the evolution of the number of un-absorbed 
primary $\gamma$-rays, i.e. without the contribution of the 
secondary (cascade)  component. 
} 
\label{cas_dev2} 
\end{center}
\end{figure}

The idea behind the  attempt to interpret TeV 
radiation  of Mkn~501 by the 
intergalactic cascade is the following.  The number of photons in  a given 
energy band of the cascade spectrum may  essentially exceed 
the number of $\gamma$-rays in the absorbed  primary $\gamma$-ray  
spectrum.  The $\gamma$-rays of highest  energies,
during their propagation trough the intergalactic photon fields,   
initiate cascades, which  produce many  lower energy photons, i.e. 
transfer the energy from the primary highest energy photons to
lower energy $\gamma$-rays.   Thus,  the 
cascade  somewhat ''move'' the source closer to the observer. 
In order to illustrate this effect, in 
Figs.~\ref{cas_dev1} and ~\ref{cas_dev2}
we show the  number of cascade $\gamma$-rays  
in  5 energy bands as a function of the penetration depth.  
In the case of  a cascade initiated by a 
monoenergetic primary $\gamma$-ray  of energy $E_0$ 
(Fig.~\ref{cas_dev1}),  
the number of photons  in all energy bands 
sharply  increases and reaches its maximum at  a distance 
$R_\ast$ which  is determined approximately by  the condition
$\tau(E_0) \sim 1$ or $R_\ast \simeq L(E_0)$, where $L(E_0)$
is the mean free path of primary photons.   
Beyond $R_\ast$ the cascade develops  more slowly. This stage 
is characterized by  a competition  between production and absorption 
processes. While the low energy (0.1 - 0.3 TeV) 
$\gamma$-rays continue to  (slowly)  grow, the number of higher energy 
(10-30 TeV) $\gamma$-rays  drops  beyond 10 Mpc,  the reason being 
the expiration  of the ``fuel'', i.e.  particles of sufficient energy,  which could support 
further  development of the cascade at these energies.  
It is interesting to note that the overall picture slightly  depends 
on energy of the primary photon  
(compare curves in Fig.~\ref{cas_dev1}  calculated for  
$E_0=10^4$ and $10^2$ TeV). This  
explains why the spectrum of $\gamma$-rays of the  
well-developed  cascade becomes  almost independent
of the primary $\gamma$-ray spectrum. 

%

The case of  the   cascade initiated by a 
broad-band spectrum of primary $\gamma$-ray photons
is more  complicated because the observer detects  
a mixture of the primary  (un-absorbed) and secondary (cascade)  
$\gamma$-rays.  
Two example  of cascades triggered by  a $E^{-2}$ type  
primary $\gamma$-ray spectrum  with 
$E_{\rm max}=10^2$  TeV (dashed lines)
and  $E_{\rm max}=10^4$ TeV (dot-dashed lines) 
are  shown in Fig.~\ref{cas_dev2}.   In order to  demonstrate  
the  photon excess caused by the cascade development,  
we show also the evolution of the number of photons calculated for 
the  simple absorption effect,  i.e. 
$J(E)=J_0 \ \exp{[-R/L(E)]}$.

\begin{figure}[tbp]
\begin{center}
\includegraphics[width=0.9\linewidth]{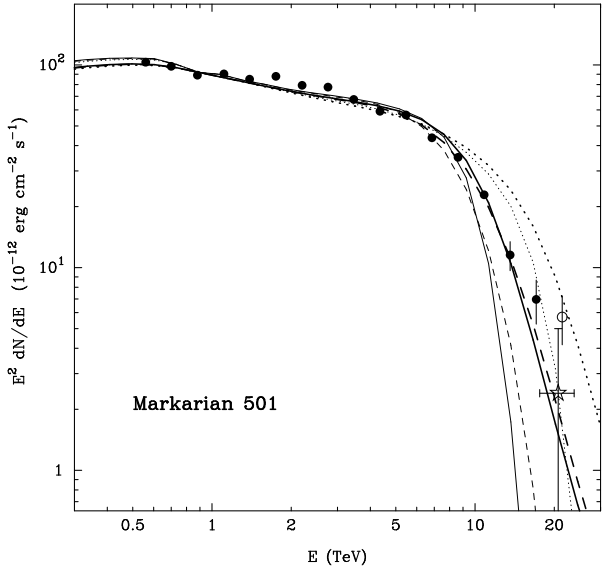}
\caption{Intergalactic cascade radiation  
spectra  initiated  by primary $\gamma$-rays from 
Mkn~501. The solid, dotted, and dashed  lines correspond to 
the Model I, Model II, and Model III of CIB, respectively. 
The light and heavy lines  represent  the 
predicted cascade $\gamma$-ray fluxes before and after 
convolution  with the Gaussian type  energy spread function 
with $20 \%$ energy resolution.} 
\label{cas_spec1} 
\end{center}
\end{figure}

In Fig.~\ref{cas_spec1} we show the  
spectra of   cascade $\gamma$-rays from  Mkn~501, calculated  
for the Hubble constant   $H_0=60 \ \rm km/s / Mpc$.   The spectrum
of the primary $\gamma$-radiation  was assumed to be power-law 
$J_0(E)\propto E^{-\alpha}$ with $\alpha=1.9$. For softer
source spectra ($\alpha>2$),   the  cascade has a small 
impact on the resulting spectrum.
For harder  primary  spectra 
with $\alpha<2$,  the most power is radiated in 
the high energy range, and therefore the resulting 
spectrum is dominated  by the ``cascade'' component. 

For a  narrow, e.g. Planckian type spectrum 
of target  photons with temperature $T_{\rm r}$, 
the spectrum of well-developed cascade  
$\gamma$-rays  has a standard power-law shape 
with photon index 1.5 at energies 
$E \ll  E^\ast \simeq  m_{\rm e}^2 c^4/ kT_{\rm r}$, 
and sharp cutoff beyond $E^\ast$. 
For a  broad-band background  photon field,  the   
spectral shape  of CIB plays  an essential  role 
in forming the cascade $\gamma$-ray spectrum, especially in the region of the cutoff. 
This can be  seen in Fig.~\ref{cas_spec1}  where the  $\gamma$-ray spectra 
are calculated for 3 different models  of CIB  presented  in Fig.~\ref{SED_CBR}.  
For  comparison, we show  also the fluxes of Mkn~501
as measured  by HEGRA  during the  high state of the source
in 1997.  The experimental points  (filled circles up to 17 TeV 
and an open circle at $\simeq 21$ TeV)  are obtained with 
an  energy resolution  of about $20 \%$.  In addition, they   contain 
$\approx 15 \%$  uncertainty on the energy scale 
throughout the entire interval  from 0.5 to 20 TeV,  and
up to factor of 2  statistical and systematic uncertainties 
in flux estimates  above 10 TeV  \citep{aharonian/::mkn501spectr:1999} 

Within these  uncertainties,  the cascade $\gamma$-ray spectrum   
corresponding to  the CIB Model II (light dotted line) agrees with 
the  HEGRA points, while the $\gamma$-ray spectra  calculated for the  
CIB Models I and III  (light solid and dashed curves,  respectively)  
pass significantly below the measured fluxes  at $E \geq  10 \ \rm TeV$
However, at these energies the spectra are very steep, therefore 
we must to take into account the limited instrumental energy resolution
when comparing the theoretical predictions with the 
differential flux measurements.  Namely,  the experimental fluxes 
should be compared  with the predictions  after {\em convolving} 
the theoretical $\gamma$-ray spectra  with the  energy resolution function.  
In Fig.~\ref{cas_spec1} the corresponding  curves, obtained assuming   
Gaussian type  instrumental energy spread function  
$G(E-E')=\exp(-(E-E')^2/2\sigma^2)/(\sqrt{2\pi}\sigma)$  with  
constant $\sigma = E/5$,   i.e. $20\%$  energy resolution, 
are shown by heavy solid (Model I), dotted (Model II) and dashed (Model III) 
lines. It is seen that after this procedure  the  $\gamma$-ray spectra 
corresponding to the CIB Models I and III quite    
satisfactorily fit  the HEGRA  points. 
At the same time   the CIB Model II 
gives  noticeably  higher $\gamma$-ray fluxes. 

\begin{figure}[tbp]
\begin{center}
\includegraphics[width=0.9\linewidth]{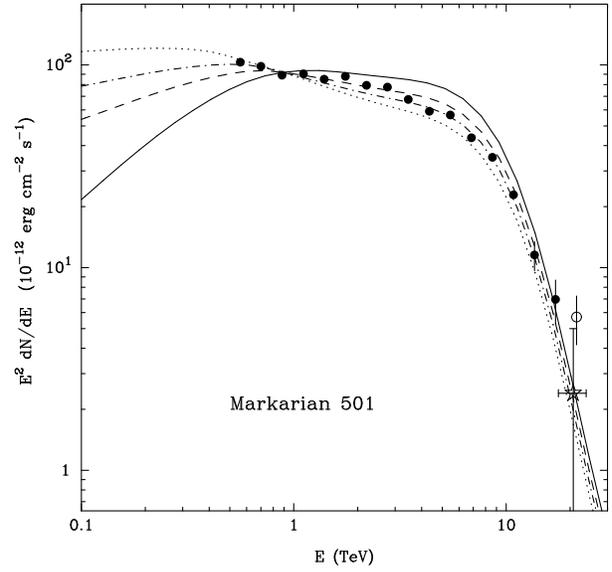}
\caption{Intergalactic cascade radiation spectra  
initiated  by primary $\gamma$-rays from 
Mkn~501. The solid, dashed, dot-dashed and dotted curves 
are the cascade $\gamma$-ray spectra 
corresponding to the the photon indices of the primary 
$\gamma$-radiation $\alpha$=1.5, 1.8, 1.9 and 2, respectively. 
The  curves  are obtained after  the 
convolution  of the cascade $\gamma$-ray spectra 
with the Gaussian type  energy spread function 
with $20 \%$ energy resolution. All curves are calculated for the CIB Model III. 
} 
\label{cas_spec2} 
\end{center}
\end{figure}  

The HEGRA stereoscopic system of telescopes has a potential to reduce the 
energy resolution down to $\sim 10 \%$ (Hofmann 2000). The recent re-analysis 
of the 1997 high state Mkn~501 data  \citep{aharonian/::mkn501newspectr:2001} 
based on  the improved energy 
reconstruction  method  confirmed  the previous flux measurements, 
except the flux upper limit  at $\simeq 21$ TeV  
which appeared by  a factor of 2.5 below  the previous flux estimate  
(the star and open circle  in  Fig.~\ref{cas_spec1},  respectively).
Therefore the $20 \%$ energy resolution used  for calculations  
presented in Fig.~\ref{cas_spec1} should be considered as 
a rather  conservative assumption. Nevertheless 
this  seems  to be,  an appropriate assumption, 
at least for the  purposes of this paper,  given the lack of 
independent  experimental  methods  for  cross-calibration of the 
energy resolution of  ground-based Cherenkov telescopes. 

At high energies the spectrum of cascade photons  weakly depends on the 
the power-law index   $\alpha$  and the maximum energy  $E_{\rm max}$
of primary $\gamma$-rays, provided that  $\alpha \leq 2$ and 
$E_{\rm max} \gg 10^2 \ \rm TeV$. In   
Fig.~\ref{cas_spec2}  four cascade $\gamma$-ray spectra are 
shown calculated  for $\alpha=1.5, 1.8, 1.9, 2$ and 
$E_{\rm max}=10^3 \, \rm TeV$ and assuming  $20 \%$ 
detector energy resolution. All spectra are calculated for the 
CIB model III.  The photon indices of primary $\gamma$-rays 
$\alpha$ = 1,8, 1.9, and 2  provide very similar     
fits for the  HEGRA points throughout the entire energy interval from 
0.5 TeV to 20 TeV.  They become distinguishable at low energies,  
but mainly because of the contributions  of primary (un-absorbed) 
$\gamma$-rays which dominate below 1 TeV. 
The very hard primary spectrum with $\alpha=1.5$ 
results in  a flatter cascade $\gamma$-ray spectrum compared to 
the measurements. On the other hand,  the primary photon index 
$\alpha$ should not exceed  2.0-2.1,  otherwise 
the efficiency of the cascade contribution becomes  very small.   
Hoverer, for the  photon index of about 2 the primary
spectrum has to  extend beyond 100 TeV in order to  provide
an adequate contribution from the cascade component.

\section{Gamma-radiation  produced by ultrarelativistic unshocked jets 
\label{sec_4}} 

The "intergalactic cascade" hypothesis requires extremely small 
intergalactic magnetic field of about $B \sim 10^{-18} \ \rm G$  
and extension of the primary $\gamma$-ray spectrum well beyond 100  TeV.
In the case of failure of any of these two conditions, the contribution of the 
secondary $\gamma$-rays produced in the intergalactic medium becomes 
negligible,  and thus the attenuation of $\gamma$-rays is reduced to the  
simple absorption effect. Consequently,   
the absorption-corrected spectrum of $\gamma$-rays given by 
Eq.~(\ref{eq:gsource})  adequately  represent the intrinsic 
TeV spectrum of the source. If so, we may face a difficulty   
with the current  common  belief that the source spectrum 
of $\gamma$-rays  should  have a ``decent''  shape, i.e. be in accord
with the predictions of conventional  astrophysical scenarios
suggested for blazars.  Indeed, none of the models of blazars, 
in general, and of Mkn~501  in particular,  allow  the  striking feature
which appears unavoidably in the absorption-corrected spectrum of 
Mkn~501 above 10 TeV (see Fig.~\ref {mkn501_60}), 
if we adopt that  the reported  high FIR fluxes are due to the truly extragalactic 
background radiation.  All versions of  both the electronic  
and  hadronic  models,  suggested so far,    
predict smooth broad-band  $\gamma$-ray spectra with  
characteristic  (gradual or sharp) steepening   above 10 TeV.

%
\begin{figure}[tb]
\begin{center}
\includegraphics[width=0.9\linewidth]{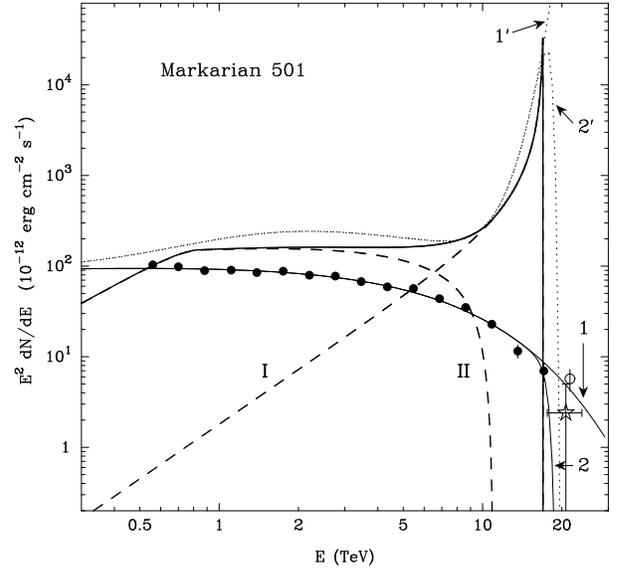}
\caption{Inverse Compton spectrum of cold unshocked 
ultrarelativistic jet with bulk motion Lorentz factor
$\Gamma = 3.33 \times 10^7$.  
The radiation component associated with comptonization of 
ambient optical photons
with narrow spectral distribution  is shown by
the dashed line I. The IC spectrum on ambient photons
with broad-band spectral distribution is shown by
the dashed line II.  
The heavy solid line represents the 
superposition of these two components.
Formally,  the dashed line II could be treated  as 
residual of the total TeV source emission 
after subtraction of the unshocked wind component I,
and therefore can be referred to the IC radiation 
of blobs in shocked jet (see Fig.~\ref{jet_scheme}).   
Fits to the observed flux of Mkn~501 are shown by thin solid lines.
Curve  1 corresponds to the fit given by Eq.(\ref{eq:hegra_501})
and curve  2 -- to the steepest possible spectrum 
above 17 TeV based on the reanalysis of Mkn~501 HEGRA data 
\citep{aharonian/::mkn501newspectr:2001}.
 Absorption-corrected spectrum of Mkn~501 for CIB
Model I is shown by dotted lines $1^\prime$ and $2^\prime$ for the 
fits to the observed spectrum 1 and 2
respectively.} 
\label{ic_0} 
\end{center}
\end{figure}  

Formally, it is possible  to  reproduce   a $\gamma$-ray spectrum 
with a sharp pile-up assuming very narrow features in  the distribution 
of accelerated particles.  It should be noticed in  this regard that  
almost all particle acceleration models predict power-law distributions
with high energy  cut-offs.  While the cutoff energy 
$E_0$ can be  estimated, generally quite confidently,  from the balance 
between  the particle  acceleration and the energy loss rates, the shape 
of the resulting particle spectrum  in the cutoff region depends on many 
specific  acceleration and energy dissipation  mechanisms.  
Remarkably,  even within the ``ordinary''   shock acceleration 
scenarios   we may expect not only spectral cutoffs, 
but also pronounced pile-ups preceding the cutoffs 
(e.g.   \citet{malkov/:1997, melrose/crouch:1997, 
protheroe/stanev:1999, drury/:1999},
see however \citet{kirk_ICRC:2001}). 
Within the proton-synchrotron  model of blazars,  such a pile-up 
would result in  a bump in the  synchrotron TeV  emission 
at  $\simeq 0.3 \eta^{-1} D_{\rm j} \, \rm TeV$  \citep{aharonian:2000},
where $\eta \geq 1$ is the so-called gyrofactor.  Thus,   the location of the 
bump  at   $E \geq$ 17 TeV should  require very large   
Doppler factor  $D_{\rm j} \geq 50  \eta$. 
In the leptonic  models the   pile-up
in the electron spectrum  would also result in  corresponding 
features in the synchrotron X-ray and inverse Compton 
TeV $\gamma$-ray spectra.        
The absorption-corrected spectra  of Mkn~501 shown in 
Fig.~\ref {mkn501_60} require very sharp pile-up in the electron spectrum. 
Whether such a  pile-up could be formed in realistic particle 
acceleration models of small (sub-pc) jets of blazars is a question of 
future  detailed studies. 

In this paper  we offer a different,  {\em non-acceleration} scenario 
which postulates  that the TeV radiation of  Mkn~501 is a result of 
comptonization of the ambient  low-frequency radiation  
by  ultrarelativistic  jet-like  outflow    with a Lorentz  factor of 
the bulk motion  of about $\Gamma \simeq (3-4) \times 10^{7}$.

The  relativistically moving plasma outflows  in forms of    
jets or winds,  are  common for many  astrophysical phenomena  
on both galactic  or extragalactic scales  
\citep[see e.g.][]{mirabel:2000}.  
Independent of  the origin of these  relativistic outflows, the concept of 
the jet  seems to be the only successful approach to understand  the complex 
features  of  nonthermal radiation of blazars,  microquasars  and 
GRBs. The Lorentz factor of such outflows could be extremely 
large. In particular,  in the  Crab Nebula the Lorentz-factor of the 
MHD wind  is estimated   between  $10^6$ and $10^7$ 
\citep{rees/gunn:1974, kundt/krotscheck:1980, 
kennel/coroniti::confinement:1984}.
The  conventional  Lorentz-factors of jets  in 
the inverse Compton models of $\gamma$-ray blazars,
are  rather  modest,  $\Gamma \sim 10$. However  there are no apparent 
theoretical  or observation arguments against the bulk motion 
with extreme Lorentz-factors  (see e.g. discussion by 
\citet{celotti/::mkn421:1998} on  Mkn~421). 
\citet{meszaros/rees::jets:1997} have
shown that  in the context of cosmological GRBs  
the magnetically dominated jet-like  outflows from stellar mass 
black holes may attain extreme  bulk Lorentz factors exceeding 
$10^6$.  We are not aware  of similar calculations (and conclusions) 
concerning the massive black holes - the engines of  AGN.
Therefore we will limit our discussion by postulating existence of 
an  outflow in Mkn~501 which moves, at least  at the initial  stages of 
its  propagation  in the vicinity of the central black hole,  with an extreme 
bulk Lorentz  factor $\geq 10^7$.  If true, it is almost obvious that the outflow 
should have MHD origin, the energy being extracted from the rotating 
black hole  through e.g. the Blandford-Znajek type mechanism 
(see \citet{blandford:2000} for a recent review). 
In proximity of  the accretion disk,   the outflow 
should be Poynting-flux  dominated in order to avoid the 
Compton drag,  and only    at relatively   
large  distances from the central object,  
where the photon density is significantly  reduced,   
an essential  fraction of the  electromagnetic energy  must be
transfered to  the kinetic energy of bulk motion.

From a  cold relativistic  outflow  we do  not expect significant 
synchrotron radiation. Indeed,  although the energy of  electrons in the frame 
of observer can be as large  as 10 TeV,   
they move  together with  magnetic field, and thus they do not 
emit synchrotron photons. Nevertheless,   the cold ultrarelativistic 
outflow  could be  {\rm visible} through the  
inverse Compton  $\gamma$-radiation of wind electrons. 
Apparently,  in this scenario  
the dense target  photon field is the  
second important ingredient  for effective production of $\gamma$-rays.    
Remarkably, even in the case  of relatively weak BL Lac objects  
there could be  several important sources of  infrared and optical 
emission  within the inner sub-parsec region of the 
central source --  IR emission from the dust torus, broad-line 
emission from fast moving clouds, starlight, etc. It is easy to 
show that the Compton optical depth  $\tau_{\rm C}$ in this 
region could be  as large as 1 \citep[see e.g.]{celotti/::mkn421:1998}. 
In fact,   we have to introduce  some additional
conditions to avoid the destruction of the jet due to the Compton drag. 
On the other hand, since for  $\tau_{\rm C} \leq 1$ the 
$\gamma$-ray luminosity is proportional to the product 
$L_{\rm j} \times \tau_{\rm C}$, the optical depth should  be  
close to  1 in order to avoid  assumptions which would 
require very large  jet power $L_{\rm j}$.
Obviously,  the most favorable  value for $\tau_{\rm C} \leq 1$  
lies   between 0.1 and 1.  Because of  extremely large  
bulk Lorentz factor $\Gamma \geq 10^7$, the Compton scattering on the 
ambient optical photons and near-infrared photons  proceeds in deep 
Klein-Nishina regime. Therefore the resulting $\gamma$-radiation  should have
a  very narrow distribution with a maximum at the edge of the spectrum
$E \approx E_{\rm e}= m_{\rm e} c^2 \Gamma$.  
Meanwhile,   the IC scattering  on  low-frequency target photon components, 
e.g. on far infrared  photons of the dust torus,  would still take place in the 
Thomson regime, and thus would produce  a smooth broad-band  
$\gamma$-ray spectrum.

%
\begin{figure}[tb]
\begin{center}
\includegraphics[width=0.9\linewidth]{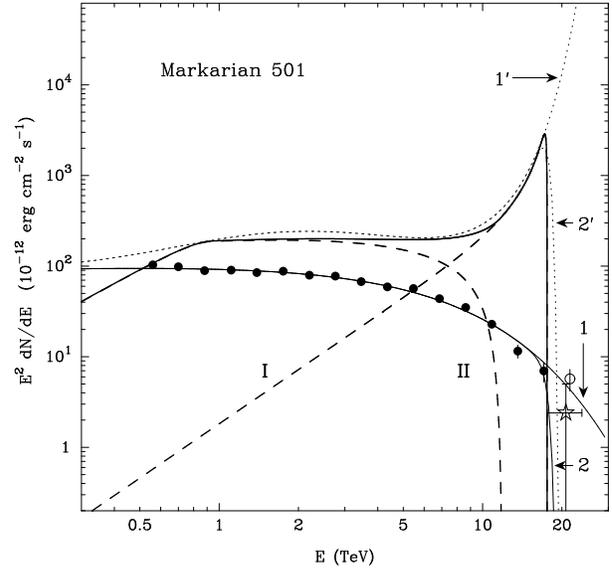}
\caption{The same  as Fig.\ref{ic_0}, but for CIB Model III.}
 
\label{ic_1} 
\end{center}
\end{figure}  

In Figs.~\ref{ic_0} and \ref{ic_1} we demonstrate that
the overall absorption-corrected spectrum of 
Mkn~501 can be satisfactorily interpreted in the terms of 
bulk inverse Compton emission of the jet,   assuming a 
specific  ambient radiation field consisting of two - narrow 
(Planckian)  type radiation with temperature 
$kT \sim 1 \, \rm eV$ (curve I)  and broader IR 
radiation, which formally could be presented in a 
power-law form $n(\epsilon) \propto \epsilon^{-m}$  (curve II).
More specifically, the best fit for the CIB Model I 
is achieved for $k T = 2$~eV and $m=1.8$,  with the  ratio of the 
energy densities of two components of about $r=w_1/w_2=163$. 
The CIB Model III requires similar, although  slightly different 
parameters:    $k T = 0.2$~eV, $m=1.8$, and $r=9.5$. 
In both cases  the bulk Lorentz factor 
$\Gamma=3.33 \times 10^7$  is assumed.

\begin{figure*}
\centering
\includegraphics[width=16cm]{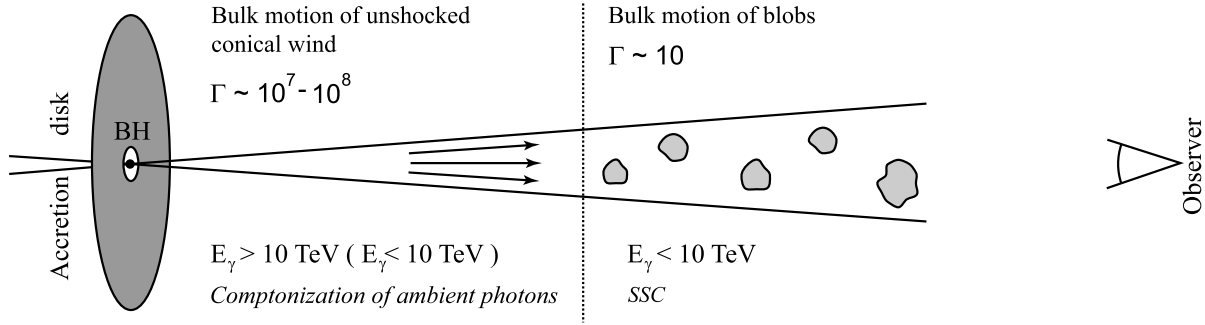}
\caption{%
Sketch of the two-stage $\gamma$-ray production scenario 
in Mkn~501. At the first stage,  the observed highest energy
$\gamma$-rays above 10 TeV with a 
sharp spectral form  (``pile-up'') 
are produced due to comptonization of the ambient 
optical photons of energy $\sim 1-2 \ \rm eV$ 
by the cold  ultrarelativistic jet with bulk motion Lorentz factor 
$\Gamma \sim 10^7-10^8$. Low energy $\gamma$-rays, 
$E_\gamma  \leq 10 \ \rm TeV$ can be also produced due to comptonization 
of far infrared photons with broad spectral distribution. 
At the second stage,   relatively low energy ($E_\gamma  \leq 10 \ \rm TeV$)
$\gamma$-rays are produced  in the blobs moving with Lorentz factor 
of about $\Gamma \sim 10$ in accordance with 
the standard   SSC or external IC models. 
} 
\label{jet_scheme} 
\end{figure*}

The ``Klein-Nishina''   component, caused by inverse Compton 
scattering on the ``hot''  narrow photon distribution,  
is of prime interest because this component provides the 
most critical part of the spectrum -
the pile-up.  Moreover, we can omit the second (``Thomson'') 
component of $\gamma$-radiation   (for example  assuming larger 
values for the  ratio $r=w_1/w_2$ compared to  the ratios used in  
Figs.~\ref{ic_0} and ~\ref{ic_1}). This should  not be a  
problem  for the  model, because   we may compensate this loss  by  
assuming that the low  ($E \leq 10 \, \rm TeV$) energy part of the  
$\gamma$-ray spectrum  is formed  at later  stages of evolution 
of the jet, namely  due to transformation of a fraction of the  
bulk kinetic  energy of the  jet into a random energy of relativistic electrons, 
e.g in the form of so-called blobs (plasmons) moving with Lorentz factor 
of about 10.  This can be realized through  the acceleration of electrons by    
the terminal shock  which may be created  
at interaction of the jet with the ambient matter
(here again,  we could see an interesting analogy with the 
Crab Nebula).  Actually this is the stage which is  treated
in all conventional, SSC or external Compton   models of 
X-ray and  $\gamma$-ray emission of TeV blazars 
\citep{sikora/madejski:2001} with ``standard'' 
jet parameters like Doppler factor of about 10-30, and 
the magnetic field of about  0.1-1 Gauss 
\citep{inoue/takahara:96,bednarek/protheroe:97,
tavecchio/:1998,mastichiadis/kirk:97,krawczynski/coppi/:2000}.
The two-stage scenario of jet radiation  is illustrated 
in Fig.~\ref{jet_scheme}.

\section{Discussion}

It is believed that  the diffuse extragalactic infrared background radiation 
may have  a  dramatic  impact on the models of  TeV blazars. In particular,   
the recently reported high CIB  fluxes,  both at NIR and FIR 
bands, imply  that we  detect significantly absorbed TeV radiation
even  from relatively  nearby   objects like Mkn~421 and Mkn~501. 
Due to the energy-dependent mean free path of 
$\gamma$-rays in the intergalactic medium, the detected spectra 
of TeV emission from extragalactic  objects  at  cosmological 
distances  significantly  deviates  from  the source spectra. 
The intergalactic absorption does not simply imply spectral
cutoffs, but rather  modulates the  primary spectrum  
from sub-TeV to multi-TeV energies. For some specific  CIB  models, 
the ``high-state'' TeV spectrum of Mkn~501 after correction 
for the intergalactic absorption  becomes 
rather flat  (Fig.~\ref{mkn501_60} ), 
${\rm d}N/{\rm d} E \propto E^{-2}$  
\citep[e.g.][]{coppi/aharonian::SSC:1999,konopelko/:1999} 
or even harder  
\citep{guy/:2000,renault/:2001,jager/stecker:2001}, if the 
CIB density  of the mid-infrared ``valley''   
exceeds  $\nu F_\nu \geq 4 \  \rm nW/m^2 sr$ 
\citep{aharonian/::mkn501spectr:1999}.     
Although formally such spectra can be described  within the 
conventional SSC models, the detailed treatment of the 
problem within the one-zone SSC model requires 
quite uncomfortable  jet parameters, in particular very
large  Doppler  factor $\delta_{\rm j}  \geq 100$,
and  very weak   magnetic field  
$B \leq 0.01 \ \rm G$. 
If so, this  perhaps would require a serious consideration  
alternative models for TeV emission like  the proton-synchrotron 
model which can satisfactorily  fit the data of 
Mkn~501 and Mkn~421 even for quite high CIB fluxes
(Aharonian 2000).  

However all these models fail to explain the sharp pile-up which 
appears at the end of the  ``reconstructed''  spectrum 
of Mkn~501, if the reported FIR  fluxes correctly describe the level of the 
diffuse cosmic background radiation. 
Remarkably, such a pile-up is  unavoidable 
not only due to  the reported extremely high flux at 
$60 \mu \rm m$, which obviously needs further confirmation,
but also,  albeit in a less distinct form,  due to the fluxes at longer, 
$\lambda \geq 100 \ \mu \rm m$ wavelengths reported 
by several independent groups. To avoid  such an unusual shape 
of $\gamma$-radiation, recently  several dramatic  assumptions  
have been proposed concerning  the nature of the detected signals
or the violation of the Lorenz-invariance at very high energies.   

The main objective of this work was the attempt to  investigate  other
possible ways to overcome the   ``IR background -- TeV gamma-ray crisis''.
In particular, we studied two possibilities:  (i) assuming that 
the observed TeV spectrum of $\gamma$-rays from Mkn~501 is 
formed due to the  
cascade initiated by primary $\gamma$-rays in the 
intergalactic medium, and (ii) assuming that the 
detected highest energy $\gamma$-rays
are result of {\it bulk-motion comptonization} of the
ambient optical photons by the 
ultrarelativistic unshocked conical wind emerging from the central source.

\subsection{Rectilinear intergalactic cascade radiation}

The electromagnetic  cascades  in the extragalactic photon fields 
result  in a  non-negligible  increase of the 
effective {\em propagation}  path of $\gamma$-rays.
Remarkably,  the results presented in Sec.~3 show that 
the cascade $\gamma$-ray spectra can quite well  
fit the time-averaged high state spectrum  of Mkn~501.
An important  feature of the well-developed cascade radiation is its  
standard  spectral shape which  is determined by the 
propagation effects and weakly  depends on the details 
of the primary radiation of the central source.  
Thus,  this hypothesis could give a natural explanation 
for  the  quite stable spectral shape 
of the source in high state, despite 
dramatic variation of the  absolute flux on timescales less 
than several hours   \citep{aharonian/::mkn501spectr:1999, 
aharonian/::mkn501timing:1999}.   Note that the  
spectral change during the  strong  April 16, 1997  flare reported by 
the CAT collaboration  \citep{djannati-atai/:1999},
as well as the noticeable 
steepening   of the spectrum in a low state of Mkn~501, 
found by the HEGRA collaboration 
\citep{sambruna/:2000, aharonian/::mkn501lwst:2001} 
do not contradict to this statement.  These effects, in fact,  could be 
caused by  variations  of the ratio of  the ``cascade''
component to the overall (``unabsorbed'' plus ``cascade'') 
flux of $\gamma$-rays  due to, for example,    
reduction  of the  maximum energy in the primary $\gamma$-ray 
spectrum  at the low or quiescent states.

The ``intergalactic cascade'' hypothesis requires extremely small,
at least in the direction of Mkn~501,  intergalactic magnetic fields,
in order to avoid   the significant  time delays.  
Even a tiny  intergalactic magnetic field
of about $10^{-18} \, \rm G$ should  lead to delays  of 
arrival of  TeV $\gamma$-rays, compared to the  associated 
low energy photons,   by more than several days. This obviously  
contradicts  the observed X/TeV  correlations
\citep[e.g.][]{pian/:1998, catanese/:1997, krawczynski/coppi/:2000}.
Although indeed  very speculative,  formally such small fields  
cannot be excluded,  especially if we  take  into account 
that the typical scale  of the  so-called intergalactic voids, 
where the magnetic field could be arbitrarily   small,  is estimated 
as  $\sim 120 (H_0/100 \ \rm km /s  / Mpc)^{-1} \, \rm Mpc$ 
\citep{einasto:2000},  i.e.  quite comparable  to  the distance to Mkn~501.   
  
The second condition  for realization of this model -- the extension of 
the primary spectrum of Mkn~501 out to 100 TeV --  also is a  
very robust requirement. Although the discussion of possible 
mechanisms of production of such energetic photons
is beyond the framework of this paper, we note that  
the condition  of $E_{\rm max} \gg 100 \ \rm TeV$  cannot be easily 
accommodated by the  conventional  (especially, leptonic ) models of 
high  energy radiation of Mkn~501.

Apparently, the ``intergalactic cascade''  hypothesis  
which simply postulates very low intergalactic magnetic fields and 
very energetic  primary $\gamma$-radiation of the source,
cannot provide conclusive predictions for  the time 
correlations of TeV $\gamma$-rays  
with nonthermal radiation components at other energy bands.
At the same time, this hypothesis   allows robust  calculations of 
spectral characteristics of $\gamma$-radiation at different 
depths of the cascade  development. 
Therefore the most  straightforward (and perhaps relatively easy) 
inspection of this  hypothesis would be  the study  of 
spectral  characteristics of  TeV $\gamma$-rays  detected from   
sources with different redshifts $z$.

\subsection{Inverse Compton radiation of  ultrarelativistic cold outflow}  
 
Although the assumption about the existence of  a cold ultrarelativistic
outflow  with an extreme  Lorentz factor of bulk motion
$\Gamma \sim 3.5 \times 10^7$ is  somewhat  unusual 
and perhaps even  provocative (at least it has not been 
discussed in the literature before),  it cannot be  {\it a priori}  ruled out.
A  similar scenario most probably  takes place, although
on  significantly smaller scales, in  environments of pulsars.
The rotation-powered pulsars eject plasma in the form 
of relativistic winds which carry off bulk of the rotational energy.
At a distance $d \leq 1 \, \rm pc$ the wind is terminated by a strong standing
reverse shock which accelerates particles and randomizes their pitch angles. 
This results in formation of  strong synchrotron and inverse Compton
nebulae.  On the other hand,  it is generally believed  that  the region between 
the pulsar magnetosphere and the shock   is invisible because  
the electrons move together with magnetic field and thus do not emit synchrotron 
radiation. However,  recently it has been argued  that such  
winds could be  directly observed through their  inverse Compton emission,
the low-frequency seed photons for comptonization being provided by the 
neutron star in the case of radiopulsars like Crab or Vela  
\citep{bogovalov/aharonian:2000}
or by the optical companion star in the case of 
binary pulsars 
\citep{kirk/ball:2000}.    If such a cold 
ultrarelativistic conical wind  can  indeed be produced in the proximity of the 
central rotating black hole, 
because of  possible existence of dense photons fields  in the 
inner sub-parsec region, it 
could be a very powerful  emitter  of inverse 
Compton $\gamma$-radiation.  The later  however would not  be accompanied 
by noticeable synchrotron radiation. 

The possibility to disentangle  the multi-TeV emission with a characteristic 
sharp pile-up at the very end of the spectrum, 
$E_\gamma \simeq m_{\rm e} c^2 \Gamma$,    
associated with the unshocked jet, from the sub-10~TeV emission 
associated with the shocked structures (e.g. blobs) in the jet (see
Fig.~\ref{jet_scheme}) ,   
not only may solve the  ``IR background -- TeV gamma-ray crisis''   but   also 
would allow  more relaxed parameter space for  interpretation 
of X-rays and the  {\it remaining} low energy ($\leq 10 \ \rm TeV$)
$\gamma$-rays   within the conventional  SSC  scenario. 
Consequently,  this  offers  more
options for interpretation of X-ray/TeV $\gamma$-ray correlations both on 
small  ($t \leq $ several hours)  and large (weeks or more) timescales.  
If the overall TeV radiation of Mkn~501 indeed consists 
of two, {\it unshocked} and {\it shocked}  jet radiation components
(curves I and II in Figs.~\ref{ic_0},~\ref{ic_1}), 
we may expect essentially different time  
behaviors of these radiation components. In particular, the ``unshocked jet'' 
$\geq 10 \, \rm TeV$  radiation should arrive earlier than the 
SSC components of radiation consisting of synchrotron X-rays and 
sub-10 TeV $\gamma$-rays. Generally,  all principal parameters 
of blobs like the magnetic field $B$,  the 
maximum energy of electrons $E_{\rm max}$, 
the radius of the blob, {\em etc.}, which define  the spectra and absolute 
fluxes of synchrotron X-rays and IC TeV $\gamma$-rays, may dramatically 
evolve in time \citep{coppi/aharonian::SSC:1999}.  Therefore  we should expect
strong X/TeV correlations, especially 
during strong flares of the source. These correlations may be realized in 
quite different forms, i.e. it  could vary from flare to flare  
depending on the time-evolution of specific parameters of blobs.    
At the same time the IC radiation of the unshocked jet 
significantly depends  only on the Lorentz 
factor of the jet $\Gamma$. 
If the latter during short flares remains unchanged, we should expect 
rather stable spectral shape of the $\geq 10 \ \rm TeV$ emission, 
despite strong  variations of the SSC radiation (X- and $\gamma$-ray)  
components.  However, it should be  noticed that because of possible external broad-band
infrared  radiation fields, the unshocked jet radiation component may
significantly contribute  
to the low energy TeV radiation as well (see Sect.~\ref{sec_4}). 
If so,  it would enhance correlations between high and low energy bands of TeV
radiation,  On the other hand,  it would reduce and make 
more complicated correlations between the low energy TeV $\gamma$-rays
and synchrotron X-rays.  
   
On  larger time-scales, e.g. in a low state of the source which could
last  weeks or months,  both the particle density   
and the Lorentz factor of the 
unshocked jet  may be smaller than in high states. This would  not only reduce the  
luminosity of all components of nonthermal radiation, but also would   
make softer the overall $\gamma$-ray spectrum. Apparently, 
the pile-up associated with the radiation of cold wind would be 
shifted towards lower energies.  Although its intensity could be  
significantly  weakened,  nevertheless the search for a relevant spectral feature   
seems to be an interesting opportunity,  especially if we take into 
account that   at these energies the intergalactic absorption effect also 
becomes weaker.
Because of  low photon statistics such studies can 
be effectively performed only by next generation ground-based detectors.

The straightforward proof  of the suggested model 
would be {\em direct} detection of TeV $\gamma$-ray spectra 
with characteristic sharp high energy pile-up.  Apparently 
this requires  nearby  TeV sources   with $d \ll$ 100 Mpc, 
thus  the main  fraction of  $\geq 10 \ \rm TeV$ $\gamma$-rays  would 
arrive without significant intergalactic 
absorption.   Unfortunately,   all currently known BL Lacs  are located  
beyond  100 Mpc. This significantly limits the chances for direct detection of 
sharp pile-ups in the spectra of BL Lacs or other type of blazars,  
unless the initial Lorentz factors of ultrarelativistic outflows  
in some objects do  not exceed  $\Gamma \leq 10^7$.    

Important tests of the suggested two stage ({\it pre-shock} 
plus {\it post shock})  scenario of the TeV radiation of jets can be 
provided also by  the search 
for correlations (or lack of such correlations) of high energy (multi-TeV) 
$\gamma$-radiation with both the low energy (e.g. 1-3 TeV)    
$\gamma$-rays and synchrotron X-rays. 
The low statistics of  (heavily absorbed) $\gamma$-rays above 10 TeV 
makes the search for such correlations rather difficult, and requires 
ground-based instruments with very large, $\gg 10^5 \, \rm m^2$ 
detection areas in this energy domain.  The new generation imaging Cherenkov
telescope arrays like CANGAROO-3, H.E.S.S. and VERITAS should be able,
hopefully,  to perform  such correlation studies.

\begin{acknowledgements}
We thank Paolo Coppi, Julien  Guy and Henric 
Krawczynski for fruitful discussions. AP and AT thank 
Max-Planck-Institut f\"ur Kernphysik  for its generous hospitality.
We are very grateful the anonymous referee for her/his useful
suggestions which significantly improved the paper.
In particular, the referee called our attention to the fact that
the convolution of the predicted cascade spectrum with the
instrumental energy spread function should improve the
agreement with the observed TeV spectrum of Mkn~501.
\end{acknowledgements}

\bibliographystyle{myapj}
\bibliography{cib}

\end{document}